\documentclass[journal]{IEEEtran}

\usepackage{mathrsfs}
\usepackage{latexsym}
\usepackage{graphicx}
\usepackage{epsfig}
\usepackage{epstopdf}
\usepackage{subfigure}
\usepackage{array}
\usepackage{amsmath}
\usepackage{amssymb}
\usepackage{color, soul}
\usepackage{amsthm}
\usepackage{enumerate}
\usepackage{algpseudocode}
\usepackage{algorithm}
\usepackage{bm}
\usepackage{balance}

\theoremstyle{plain} 

\theoremstyle{definition}

\def\bal#1\eal{\begin{align}#1\end{align}}

\newcommand{\bbeta} {\boldsymbol{\beta}}
\newcommand{\bmu} {\boldsymbol{\mu}}
\newcommand{\bxi} {\boldsymbol{\xi}}

\newcommand{\bH}{{\bf H}}

\newcommand{\bn}{{\bf n}}

\newcommand{\bR}{{\bf R}}

\newcommand{\bI}{{\bf I}}
\newcommand{\bU}{{\bf U}}

\newcommand{\bA}{{\bf A}}

\newcommand{\bQ}{{\bf Q}}

\newcommand{\ba}{{\bf a}}
\newcommand{\bb}{{\bf b}}
\newcommand{\bh}{{\bf h}}

\newcommand{\bd}{{\bf d}}
\newcommand{\bq}{{\bf q}}

\newcommand{\bz}{{\bf z}}
\newcommand{\bo}{{\bf 0}}

\newcommand{\bw}{{\bf w}}

\newcommand{\bp} {\begin{proof}}
\newcommand{\ep} {\end{proof}}

\newcommand{{\Rb}} {\right)}

\newcommand{{\Rf}} {\right\}}
\begin{document}

\title{Intelligent Reflecting Surfaces Assisted Secure Transmission Without Eavesdropper's CSI\thanks{
The authors are with the  School of Information and Communication Engineering, Xi'an Jiaotong University,  and also with the Ministry of Education Key Lab for Intelligent Networks and Network Security, Xi'an Jiaotong University, Xi'an 710049, China (e-mail:  xjbswhm@gmail.com; dlm$\_$nwpu@hotmail.com;bjl19970954@stu.xjtu.edu.cn)}}

\author{Hui-Ming Wang, \emph{Senior Member, IEEE}, Jiale Bai, and Limeng Dong}

\maketitle

 \pagenumbering{gobble}

\begin{abstract}
In this letter, improving the security of an intelligent reflecting surface (IRS) assisted multiple-input single-output (MISO) communication system is studied. Different from the ideal assumption in existing literatures that full eavesdropper's (Eve's) channel state information (CSI) is available, we consider a more practical scenario without Eve's CSI. To enhance the security of this system given a total transmit power at transmitter (Alice), we propose a joint beamforming and jamming approach, in which a minimum transmit power is firstly optimized at Alice so as to meet the quality of service (QoS) at legitimate user (Bob), and then artificial noise (AN) is emitted to jam the eavesdropper by using the residual power at Alice. Two efficient algorithms exploiting oblique manifold (OM) and minorization-maximization (MM) algorithms, respectively, are developed for solving the resulting non-convex optimization problem. Simulation results have been provided to validate the performance and convergence of the proposed algorithms.
\end{abstract}

\begin{IEEEkeywords}
physical layer security, intelligent reflecting surface, secrecy rate,  oblique manifold
\end{IEEEkeywords}

\section{Introduction}

Intelligent reflecting surface (IRS) is proposed as a promising energy-efficient and cost-effective technology for reconfiguring wireless propagation environment via software-controlled reflection. Specifically, IRS is a planar surface comprising a large number of low-cost passive reflecting elements, each of which is able to change the phase for the incident signal independently, thereby collaboratively achieving passive reflect beamforming. IRS has been recognized as a strong candidate for the future wireless network \cite{Wu-02}.

Motivated by these advantages, IRS is recently combined with physical layer security (PLS) to deal with secure communication. By adjusting phase shifts coefficients at IRS, the signals reflected by IRS can add constructively with those from direct path to enhance the desired signal power at the legitimate user (Bob), and destructively with those from direct path to reduce the signal power at the eavesdropper (Eve). As a consequence, Bob's signal-to-noise ratio (SNR) is increased while Eve's is decreased and hence a higher secrecy rate can be achieved. Several algorithms were established to maximize the secrecy rates of IRS-assisted MISO transmission, including single user case \cite{Shen-19}-\cite{Guan-19} and multi-user downlink case \cite{Chen-19}. In \cite{Dong-19}, the secrecy rate of an IRS-assisted MIMO wiretap channel was studied for the first time. All these studies indicate that by jointly optimizing the active transmit beamforming at the transmitter (Alice) and passive reflect beamforming at IRS, Bob's secrecy performance can be greatly enhanced.

However, all these aforementioned works are simply based on an ideal assumption that Eve's channel state information (CSI) is perfectly known, which is not practical since Eve is usually hidden and passive who does not actively exchange CSI with Alice. Therefore, all the proposed secrecy schemes in  \cite{Shen-19}-\cite{Dong-19} are invalid in the practical case. Motivated by this, in this letter, we investigate IRS-assisted secrecy transmission without eavesdropper's CSI. The main innovations and contributions are in three aspects:

1) We propose a joint beamforming and jamming scheme to enhance security without eavesdropper's CSI, where we minimize the  power of confidential signal to meet the quality of service (QoS) at Bob and allocate all residual power to transmit artificial noise (AN) to jam the eavesdropper.

2) We propose an oblique manifold (OM) algorithm to solve the non-convex optimization problem. A minorization-maximization (MM) algorithm is also investigated.

3) Compared to the full CSI case, security could still be guaranteed by increasing of QoS threshold at Bob as well as the number of reflecting elements at IRS. The performance of OM algorithm is better than of the MM algorithm.

\emph{Notations}: 
For a vector $\ba=[a_1,\cdots,a_n]$, $||\ba||$ denotes the Euclidian norm,
$\emph{diag}(\ba)$ denotes a diagonal matrix whose entries are $a_1,\cdot\cdot\cdot,a_n$, and $\emph{unt}(\ba)=\left[\frac{a_1}{|a_1|}, \frac{a_2}{|a_2|}, \cdots, \frac{a_n}{|a_n|}\right]^T$; $\lambda_{max}(\bA)$ and $\emph{tr}(\bA)$ denote maximum eigenvalue and trace of matrix $\bA$, respectively; $\emph{arg}(a)$ and  $\Re\left \{ a\right \}$ denotes phase and real part of $\emph{a}$; $\mathbb{E}\left \{ \cdot \right \}$ denotes statistical expectation;   $\circ$ denotes Hadamard product between two matrices.
\section{System Model}
Consider an IRS-assisted communication system, which consists of Alice, Bob, Eve, and an IRS shown in Fig. 1. We assume that Alice is equipped with $N_t$ antennas, both Bob and Eve are equipped with a single antenna, and the IRS has $L$ reflecting elements. We consider a quasi-static flat-fading channel model. Without eavesdropper's CSI, we propose a joint beamforming and jamming scheme where Alice sends both information and AN signals concurrently. The received signals $y_B$, $y_E$ at Bob and Eve can be expressed as
\bal
\label{eq.ch}
y_B = (\bh_{IB}^H\bQ\bH_{AI}+\bh_{AB}^H)(\bw x+\bn_a)+\bxi_B \ ,\\
\quad y_E = (\bh_{IE}^H\bQ\bH_{AI} +\bh_{AE}^H)(\bw x+\bn_a)+\bxi_E \ ,
\eal
where  $\bw,\bn_a\in\mathbb{C}^{N_t\times1}$ represent the transmit beamformer and AN at Alice, $\bQ \triangleq \emph{diag}(e^{j\theta_1},e^{j\theta_2},\cdots,e^{j\theta_L})$ is the phase shift matrix, $\theta_i$ is the phase shift of \emph{i}-th reflecting element, $\bH_{AI}\in\mathbb{C}^{L\times N_t}, \bh_{IB}\in\mathbb{C}^{L\times 1}, \bh_{IE}\in\mathbb{C}^{L\times1}, \bh_{AB}\in\mathbb{C}^{N_t\times1}$ and $\bh_{AE}\in\mathbb{C}^{N_t\times1}$ represent channels from Alice to IRS, IRS to Bob, IRS to Eve, Alice to Bob and Alice to Eve, $x$ is the transmitted signal following $\mathbb{E}\left\{|x|^2 \right\}$=1, $\bxi_B$, $\bxi_E$ are additive complex white Gaussian noises in which the entries are with zero-mean and variances ${\sigma_b}^2$ and ${\sigma_e}^2$ at Bob and Eve respectively. For briefly, we denote $\bh^H_B\triangleq\bh_{IB}^H\bQ\bH_{AI}+\bh_{AB}^H$ and $\bh_E^H \triangleq\bh_{IE}^H\bQ\bH_{AI}+\bh_{AE}^H.$  Obviously, the achievable secrecy rate in this scheme is
\bal
\label{final_rate}
\notag
C_S\triangleq\left[\log\left(1+\frac{|\bh_{B}^H\bw|^2}{{\sigma_b}^2+\bh_{B}^H\bR_{AN}\bh_B}\right)- \right. \\
 \left.  \qquad\log\left(1+\frac{|\bh_{E}^H\bw|^2}{{\sigma_e}^2+\bh_{E}^H\bR_{AN}\bh_E}\right)\right]^+,
\eal
where $\bR_{AN}\triangleq\mathbb{E}\left\{\bn_a\bn_a^H\right\}$. Given a total transmit power budget $P_A$ at Alice, we have 
$P_A \triangleq P_T+P_J$, where $P_T = ||\bw||^2$ and  $P_J\triangleq tr(\bR_{AN})$ are signal power and AN power, respectively.

\begin{figure}[t]
	\centerline{\includegraphics[width=2.5in]{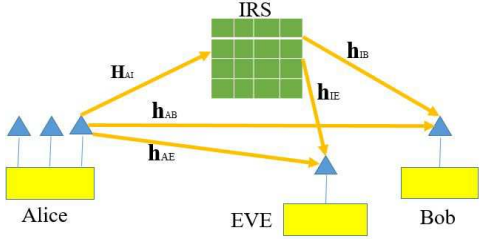}}
	\caption{An IRS-assisted MISO communication system}
\end{figure}

\section{Joint Beamforming and Jamming Scheme}
Different from the existing works in \cite{Shen-19}-\cite{Dong-19} that full CSI is available at Alice, we consider a more realistic case where although the equivalent legitimate channel $\bh_{B}^H$ is known,  both $\bh_{AE}$ and $\bh_{IE}$, i.e., $\bh_{E}^H$, are completely unknown.
Therefore, we could not directly maximize the secrecy rate \eqref{final_rate}.
To guarantee secure communication in this condition, the only solution is to \emph{increase the information rate at Bob and decrease the information leakage at Eve as much as possible}. Hence, inspired by the previous work in \cite{Wang-09}, two main procedures are included in this joint transmission scheme. Firstly, we apply AN signalling satisfying  $\bn_a\perp\bh_{B}^H$, i.e., $\bn_a$ is projected onto the null space of the equivalent channel $\bh_{B}^H$, to jam Eve only.
 Secondly, 
a minimum transmit power $P_T$ for the confidential signal $x$ is allocated to meet a target QoS constraint $\gamma$ at Bob, so that the residual power $P_J=P_A-P_T$  can be as large as possible to be used for AN to jam Eve.
We can see that both of the procedures will increase the secrecy rate as much as possible, even without knowing Eve's CSI.


Based on the above scheme, our objective is to minimize $P_T = ||\bw||^2$ by optimizing $\bw$ and $\bQ$,
subject to QoS constraints $\gamma$ at Bob, as expressed by $P1$:
\begin{equation}
P1: \underset{\bw, \bQ}{\min}\quad P_T, s.t.\quad\frac{|\left(\bh_{IB}^H\bQ\bH_{AI}+\bh_{AB}^H\right)\bw|^2}{{\sigma_b}^2}\geq \gamma, \label{QoS} 
\end{equation}
We have to note that, $\bQ$ is a diagonal matrix with each element satisfying the unit modulus constraint $|q_i|=1$ since IRS only changes the phase of the signal. The non-convex constraint (\ref{QoS}) then makes $P1$ a non-convex optimization. However, $P1$ can be equivalently transformed to a single variable optimization problem as follows. Firstly, it is obvious that the optimal solution must make \eqref{QoS} hold with equality. Secondly, for any given $\bQ$, it is known that only maximum-ratio transmission (MRT) is the optimal $\bw^{*H}$, i.e., $\bw^{*H}=\sqrt{P_T}\frac{\bh_{B}^H}{||\bh_B||}$. Therefore, by substituting $\bw^*$ into problem $P1$, one obtains that the optimal minimum transmit power is expressed as $P_T^*=\frac{\gamma{\sigma_b}^2}{||\bh_{B}||^2}$. As such, minimizing $P_T$ is equivalent to maximizing $||\bh_{B}||^2$, so the optimization problem can be expressed by $P2$
\begin{equation}
P2: \underset{\bQ}{\max}\ ||\bh_{IB}^H\bQ\bH_{AI}+\bh_{AB}^H||^2, s.t.\ |q_i|=1, \forall i, \label{QoS2}
\end{equation}
which can be further equivalently expressed as 
\begin{equation}
P3: \underset{\bq}{\min}\ \bq^H\bA\bq-\bq^H\bb-\bb^H\bq,\ 
s.t.\ |q_i|=1, \forall i,\label{unit} 
\end{equation}
where $\bq\triangleq [q_1,q_2,...,q_L]^H$, $\bb\triangleq\emph{diag}(\bh_{IB}^H)\bH_{AI}\bh_{AB}$ and $\bA\triangleq-\emph{diag}(\bh_{IB}^H)\bH_{AI}\bH_{AI}^H\emph{diag}({\bh_{IB}})$. $P3$ is still  non-convex due to nonconvex constraint \eqref{unit}. In the following, we propose two different algorithms to solve this problem.

\subsection{Oblique Manifold Algorithm}
In this subsection, we develop an Oblique Manifold (OM) algorithm to obtain a suboptimal solution of $P3$. The optimization over a manifold is locally analogous to that in Euclidean space. The unit modulus constraint is handled directly by  manifold optimization theory \cite{Absil-11}. There are some recent applications of manifold optimization in wireless communications \cite{Xu-19}, \cite{Yang-12}. In the following, we briefly introduce key steps of OM algorithm. 

Specifically, $P3$ can be equivalently transformed to the following minimization formular
\bal
P4:\ \underset{\bq}{\min}\ f(\bq)=\frac{1}{\bq^H(-\bA)\bq+\bq^H\bb+\bb^H\bq}\ s.t. \ \eqref{unit}. 
\eal
Hence, the search space of $P3$ is product of $\mathcal{M}\triangleq \left\{ \bq\in\mathbb C^{L}: |q_1|=|q_2|=\dots=|q_L|=1\right\}$ in the complex plane, which is a Riemannian submanifold of $\mathbb{C}^{L}$ with the  unit modulus constraint. To obtain the solution $\bq$, three key steps are needed in each iteration of OM algorithm.

Firstly, the tangent space of manifold $\mathcal{M}$ at the point $\bq$ is defined as the space contained all tangent vectors of manifold $\mathcal{M}$ at $\bq$ \cite{Xu-19}. The tangent space for $\mathcal{M}$ at $\bq_i$ is given by
\bal
\ {T}_{\bq_i}\mathcal{M}=\left\{\bz\in\mathbb{C}^{M}:[\bz{\bq_i}^H]_{l,l}=\bo, \forall l\in\mathcal{M} \right\} \ ,
\eal
where $\bq_i$ is the current iteration point, $\bz$ is a tangent vector at $\bq_i$. The Riemannian gradient, i.e., $grad_{\bq_i}f$, is denoted as the tangent vector with the steepest increase of the objective function, which is the orthogonal projection of the Euclidean gradient $\nabla_{\bq_i}f$ onto the tangent space ${T}_{\bq_i}\mathcal{M}$. Therefore, the Riemannian gradient at $\bq_i$ is given by
\bal
\label{gradient}
grad_{\bq_i}f&=\nabla_{\bq_i}f -\Re\left\{\nabla_{\bq_i}f\circ{\bq_i}^*\right\}\circ\bq_i \ ,\\
\notag
 f'(\bq_i) = \nabla_{\bq_i}f&=-2\frac{-\bA\bq_i+\bb}{({\bq_i}^H(-\bA)\bq_i+{\bq_i}^H\bb+\bb^H\bq_i)^2} \ ,
\eal 

Secondly, after obtaining the Riemannian gradient $grad_{\bq_i}f$, the optimization approaches designed for the Euclidean space can be transplanted to oblique manifold. For instance, we can employ conjugate gradient method with the update rule of the search direction in the Euclidean space is given by
\bal
\label{CG}
\bmu_{i+1}=-\nabla_{\bq_{i+1}}f+\alpha_i\bmu_i \ ,
\eal
Here, $\bmu_i$ denotes the search direction at $\bq_i$ and $\alpha_i$ is chosen as the Polak-Ribiere parameter \cite{Absil-11} to achieve fast convergence.
\bal
\label{parameter}
\alpha_i=max\left\{\frac{[f'(\bq_{i+1})]^T\cdot f'(\bq_{i+1})}{[f'(\bq_{i})]^T\cdot f'(\bq_{i})},0\right\} \ ,
\eal
However, $\bmu_i$ and $\bmu_{i+1}$ in \eqref{CG} lie in two different tangent spaces ${T}_{\bq_i}\mathcal{M}$ and ${T}_{\bq_{i+1}}\mathcal{M}$, so the direction cannot be directly searched. To solve this problem, an operation called \emph{transport} which maps $\bmu_i$ from tangent space ${T}_{\bq_i}\mathcal{M}$ to ${T}_{\bq_{i+1}}\mathcal{M}$ is proposed. The vector transport  is given by
\bal
\label{transport}
\mathcal{T}_{\bq_i\to\bq_{i+1}}(\bmu_i) \triangleq{T}_{\bq_i}\mathcal{M}&\mapsto{T}_{\bq_{i+1}}\mathcal{M}:\\
\notag
\bmu_i &\mapsto\bmu_i-\Re\left\{\bmu_i\circ{\bq_{i+1}}^*\right\}\circ\bq_{i+1} \ ,
\eal
Analogous to \eqref{CG}, the update rule for search direction on manifold is given by
\bal
\label{direction}
\bmu_{i+1}=-grad_{\bq_{i+1}}f+\alpha_i\mathcal{T}_{\bq_i\to\bq_{i+1}}(\bmu_i),
\eal

Thirdly, after determining the search direction $\bmu_i$ at $\bq_i$, we employ the $retraction$ to find the destination on the manifold. The \emph{retraction} for the search direction $\bmu_i$ and step size $\eta_i$ at point $\bq_i$ is given by
\bal
\label{step}
\mathcal{R}_{\bq_i}(\eta_i\bmu_i)\triangleq{T}_{\bq_{i}}\mathcal{M}\mapsto\mathcal{M}:\eta_i\bmu_i\mapsto unt(\eta_i\bmu_i) \ ,\\
\notag
\eta_i=-\eta_{i-1}\frac{[f'(\bq_i)]^T\bd_i}{[f'(\bq_i-\eta_{i-1}\bd_i)]^T\bd_i-[f'(\bq_i)]^T\bd_i}
\eal
where $\bd_i$ represents the search direction in the Euclidean space, i.e., $\bd_i=-f'(\bq_i)+\alpha_{i-1}\bd_{i-1}$.

With these key steps  introduced above, the final OM algorithm for solving $P4$ is summarized as Algorithm 1. According to \cite{Absil-11}, Algorithm 1 is guaranteed to converge to a critical point of $P3$, i.e., the point where the Riemannian gradient of the objective function is zero.

\subsection{Minorization-Maximization Algorithm}
In this subsection, we apply MM algorithm to solve $P3$. The key idea of MM is to firstly obtain an approximately upper bound of the objective function and then iteratively compute the optimal value of this bound subject to constraints. Then the converged point  is a local optimal point \cite{Dong-19}.

Specifically, let $\bq_k$ be a feasible point in current iteration satisfying \eqref{unit}, an upper bound of the objective function at the next iteration point $\bq_{k+1}$ is expressed as
\bal
\notag
&\bq_{k+1}^H\bA\bq_{k+1}-\bq_{k+1}^H\bb-\bb^H\bq_{k+1}\\
\notag
\leq&\bq_{k+1}^H\lambda_{max}(\bA)\bI\bq_{k+1}-2\Re\left\{ \bq_{k+1}^H[\lambda_{max}(\bA)\bI-\bA]\bq_{k} \right \}\\
\notag
&+ \bq_{k}^H[\lambda_{max}(\bA)\bI-\bA]\bq_{k}-2\Re\left\{ \bq_{k+1}^H\bb \right \}\\
\label{bound}
=&2L\lambda_{max}(\bA)-2\Re\left\{ \bq_{k+1}^H\bbeta \right \}-\bq_{k}^H\bA\bq_{k} \ ,
\eal
where $\bbeta\triangleq(\lambda_{max}(\bA)\bI-\bA)\bq_{k}+\bb$. Hence, after dropping the constant term in \eqref{bound}, 
$P3$ can be approximated to $P5$
\bal
P5:\quad\underset{\bq_{k+1}}{\max}\ \Re\left\{ \bq_{k+1}^H\bbeta \right \}\quad s.t. \ \eqref{unit} \ ,
\eal
Clearly, $\Re\left\{ \bq_{k+1}^H\bbeta \right \}$ is maximized only when the phases of $q_{i}$ and $\beta_i$ are equal, where $\beta_i$ is the \emph{i}-th entry of $\bbeta$. Therefore, the closed-form optimal solution to problem $P5$ is
\bal
\bq_{k+1}=\left[ e^{j arg(\beta_1)},e^{j arg(\beta_2)},\cdots,e^{j arg(\beta_L)}\right] \ ,
\eal
Then let $k = k+1$ and update $\bq_k$ iteratively until the objective function converges. By initializing a feasible point $\bq_0$ and applying MM algorithm to solve $P3$ iteratively, a local optimal solution of $\bQ$ for $P3$ can be obtained.

\begin{algorithm}[t]
	\caption{Oblique Manifold Algorithm}
	\label{alg.2}
	\begin{algorithmic}
		\State 1. Set initial point $\bq_0$ and convergence accuracy $\epsilon = 10^{-4}$;
		\State 2. Set $\bq=\bq_0$ and  $\eta_0\in (0,1)$, calculate $\bd_0$ = $\bmu_0=-\nabla_{\bq_0}f$;
		\Repeat\ \ 
		\State 3. Calculate search step size $\eta_i$ according to \eqref{step};
		\State 4. Find the next point $\bq_{i+1}$ using retraction in \eqref{step}:\\
		\quad\quad\quad$\bq_{i+1}=\mathcal{R}_{\bq_i}(\eta_i\bmu_i)=unt(\bq_i +\eta_i\bmu_i)$;
		\State 5. Determine Riemannian gradient $grad_{\bq_{i+1}}f$ in \eqref{gradient};
		\State 6. Calculate transport $\mathcal{T}_{\bq_i\to\bq_{i+1}}(\bmu_i)$ according to \eqref{transport};
		\State 7. Calculate Polak-Ribiere parameter $\alpha_i$ in \eqref{parameter};
		\State 8. Compute conjugate search direction $\bmu_{i+1}$ with \eqref{direction};
		\State 9. $i\gets i+1$
		\Until    convergence, i.e, $||grad_{\bq_i}f||_2 \leq \epsilon$
	\end{algorithmic}
\end{algorithm}

\subsection{AN Signalling Strategy}
Once $P1$ is solved, the residual power $P_J=P_A-P_T^*$ is utilized for AN signalling in the null space of $\bh_B$. Recalling (\ref{final_rate}), since the CSI of $\bh_{IE}$ and $\bh_{AE}$ is completely unknown, it is impossible to optimize the transmit covariance $\bR_{AN}$ to minimize the leakage rate. Hence, we use isotropic signalling with equal power allocation to transmit the AN signals in each dimension of $\emph{null}(\bH_B)$, where $\bH_B=\bh_{B}\bh_B^H$. $\bH_B$ is a rank-1 matrix so that the dimension of  $\emph{null}(\bH_B)$ is $N_t-1$. Hence, the transmit covariance for AN can be formulated as
 \bal
\label{optimal AN}
\bR_{AN}=\frac{P_A-P_T^*}{N_t-1}\bU_{AN}\bU_{AN}^H \ ,
\eal
where the columns in the semi-unitary matrix $\bU_{AN}$ are all $N_t-1$ eigenvectors corresponding to zero eigenvalues of  $\bH_B$. Therefore, by substituting the solution $\bw$ and $\bQ$ of $P1$ as well as \eqref{optimal AN} into \eqref{final_rate}, the final actual secrecy rate returned by this joint transmission scheme can be obtained.
\section{Simulation Result}
\begin{figure}[t]
	\centerline{\includegraphics[width=3.0in]{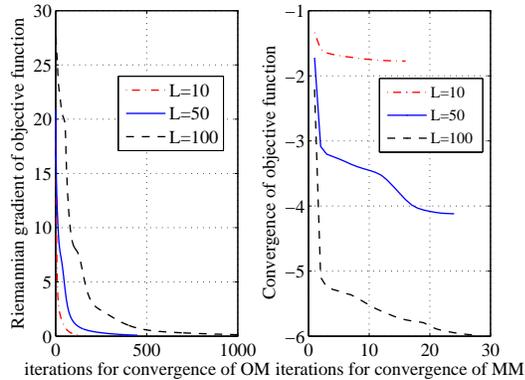}}
	\caption{ Convergence of proposed algorithms under $N_t=5$ and $\gamma=10$ dB.}
\end{figure}

We set $N_t = 5$, $P_A=5$ dBm and ${\sigma_b}^2={\sigma_e}^2=-90$ dBm. The small-scale fading of all the channels follows the Rayleigh fading and the path loss model is given by $PL=\left(PL_0-10 \rho \log_{10}(\frac{d}{d_0})\right)$ dB, where $PL_0=-30$ dB is path loss at reference distance $d_0=1$ m, $\rho$ is the path loss exponent. In the simulation we set the path loss exponents and distances of the Alice-to-IRS link, the
IRS-to-Bob link, the IRS-to-Eve link, the Alice-to-Bob link, and the Alice-to-Eve link as $\rho_{AI}=2$, $\rho_{IB}=\rho_{IE}=2.5$, $\rho_{AB}=\rho_{AE}=3$, $d_{AI}=50$ m, $d_{IB}=6$ m, $d_{IE}=7$ m, $d_{AB}=48$ m, and $d_{AE}=45$ m, respectively.

\begin{figure} 
	\centering 
	\subfigure[The achieved secrecy rate as function of $\gamma$.]{ 
		\includegraphics[width=3.0in]{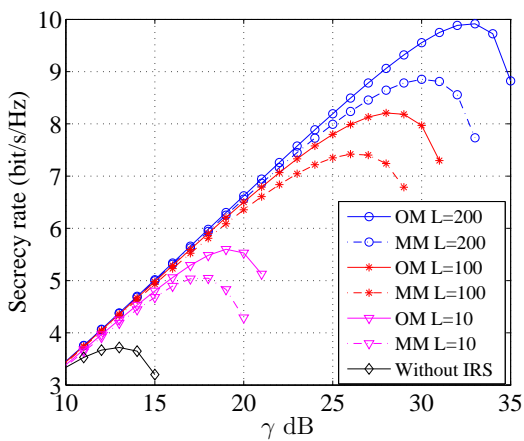} 
	} 
	\subfigure[The achieved secrecy rate for different values of $L$.]{ 
		\includegraphics[width=3.0in]{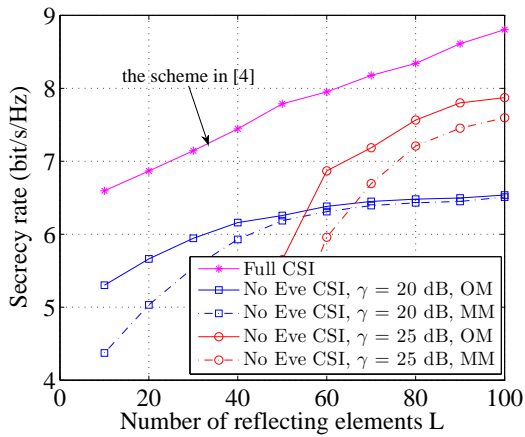} 
	} 
	\caption{ Secrecy rate under $N_t=5$. The results are averaged over 1000 randomly generated channels.} 
\end{figure}

The convergence of proposed algorithms is investigated in Fig. 2 based on a round of randomly generated channels. Note that both algorithms can guarantee convergence and MM requires significantly less iterations than OM. Furthermore, a larger $L$  requires more iterations to converge due to the reason that larger $L$ leads to larger dimensions of $\bQ$. In addition, the computational complexity of OM and MM algorithm is $\mathcal{O}\left({L^2}\right)$ for each iteration, so OM which requires more iterations needs more computational time than MM.

In Fig. 3, we can see that the QoS threshold $\gamma$ and the number of reflecting elements $L$ impact significantly on the achieved secrecy rate. In Fig. 3(a), we note security rate with IRS is significantly larger than that without IRS. It is because that IRS can supply reflected power to transmit signal, so the Alice can allocate sufficient power for AN signalling and the secrecy rate is increased. In addition, we note taht the secrecy rate achieved by OM is larger than that by MM. This is due to the fact that the optimized minimum power $P_T$ via OM is less than that via MM. Hence, more residual power can be allocated to AN signalling  by OM to jam Eve, resulting a higher secrecy rate. Furthermore, for the fixed total power $P_A$, increasing $\gamma$ can increase secrecy rate greatly because the legitimate information rate is significantly increased. However, if $\gamma$ is set to be high, no sufficient power is left for AN signalling, so that the information leakage rate dominates and the achieved secrecy rate goes down. Therefore we see the important role of the power allocation tradeoff. As $\gamma$ goes too high such that the total available power $P_A$ can not support such a $\gamma$, $P1$ becomes infeasible and the transmission fails. 

In Fig. 3(b), we note that compared to the full CSI scenario, although the secrecy rate without Eve's CSI is decreased, the security still can be guaranteed by the proposed scheme. Furthermore, for the fixed $P_A$, increasing $L$ and $\gamma$ will increase security. This is because with $L$ increasing, the IRS can provide more reflection power to transmit signal. Hence a sufficient higher $\gamma$ can be supported to exchange more information and more residual power can be allocated for AN signalling to reduce information leakage. It also implies that if the total power is limited, we can use more IRS to support a sufficient large $\gamma$ to improve the security performance.

We can see that there exists cross-point between the two curves $\gamma$ = 20 dB / 25 dB. It is because that $L$ is small, the reflected power is not sufficient. Hence if $\gamma$ is set to 25 dB, Alice needs to allocate more power for signal transmission and less power for AN singnalling, so the secrecy rate is lower than that of 20 dB. As $L$ is increasing, the reflected power is sufficient to transmit signal, so the Alice can allocate more power for AN signalling to enhance security. If $L$ is sufficient large, increasing $\gamma$ can increase the legitimate information rate. Hence when $\gamma$ is 25 dB, the secrecy rate is higher than that of 20 dB.

\section{Conclusion}
In this letter, we proposed a joint beamforming and jamming scheme to improve the PLS of an IRS-assisted  MISO system without eavesdropper' CSI. OM and MM algorithms are proposed to jointly optimize the transmit beamforming at Alice and phase shifts at IRS. Simulation results have validated the convergence of the proposed algorithms, and it is shown that the joint scheme greatly improves the security.


\end{document}